\documentclass[aps,prd,showpacs,
amsmath,amssymb,floatfix,nofootinbib,english]
{revtex4-1} 
\usepackage[T1]{fontenc}
\usepackage{textcomp}
\usepackage{amssymb}
\usepackage{epsfig}
\usepackage{graphicx}
\usepackage{esint}
\usepackage{comment}
\usepackage{color}
\PassOptionsToPackage{normalem}{ulem}
\usepackage{ulem}
\usepackage{babel}
\usepackage{slashed}
\usepackage{caption}
\usepackage{amsmath,amssymb,epsf}
\usepackage{graphicx,wrapfig,color}
\usepackage{subfig}
\usepackage{setspace}
\usepackage{makecell}
\usepackage{tabularx}
\usepackage{booktabs}
\usepackage[tablename=Table]{caption}
\usepackage{float}
\usepackage{placeins}
\usepackage{hyperref}
\usepackage[table]{xcolor}
\newcommand{\be}{\begin{equation}}
\newcommand{\ee}{\end{equation}}
\newcommand{\bea}{\begin{eqnarray}}
\newcommand{\eea}{\end{eqnarray}}

\newcommand{\ba}{\begin{array}}
\newcommand{\ea}{\end{array}}
\newcommand{\bd}{\begin{displaymath}}
\newcommand{\ed}{\end{displaymath}}

\def\gev{{\rm \,Ge\kern-0.125em V}}
\def\tev{{\rm \,Te\kern-0.125em V}}

{
{

\def\th13 {\theta_{13}}

\begin{document}

\title{Lower tensor to scalar ratio in a SUGRA motivated inflationary potential}
\author{Rathin Adhikari}
\email{radhikari@jmi.ac.in}
\author{Mayukh R. Gangopadhyay}
\email{mayukh@ctp-jamia.res.in}
\author{Yogesh}
\email{yogesh@ctp-jamia.res.in}
\affiliation{Centre for Theoretical Physics, Jamia Millia Islamia (Central University),\\ Jamia Nagar, New Delhi 110025, India.}

%\date{\today}
%%%%%%%%%%%%%%%%%%5
\begin{abstract}
A scalar potential obtained from the $D$-term in the Supergravity models, which dominates over $F$ term and is mainly responsible for the inflationary phase in the early universe, is studied. The potential with canonical kinetic terms for scalar fields in the Lagrangian, has a very slow roll feature in comparison to various other plateau type inflationary potentials. In this case, a much lower tensor-to-scalar ratio  ($r$) of  $\mathcal{O}(10^{-3})$ is achievable. The requirement of slow roll condition for the inflation potential implies that  the up type neutral scalar and the down type neutral scalar in Supergravity models are with equal field strength at the time of inflation. If this relationship holds down to the electroweak scale for the cooresponding $vev$ values of these fields, then it will indicate higher SUSY breaking scale around 100 TeV. The predicted values of the inflationary observables are well within the 1-$\sigma$ bounds of the recent constraints from {\it Planck'18} observations. The era of reheating after the inflationary phase, is also studied and the bounds on the reheating temperature ($T_{re}$) is calculated for different equation of states during reheating ($w_{re}$) for the {\it Planck'18} allowed values of the scalar spectral index ($n_s$). For our model with $w_{re}=2/3$ and $w_{re}=1$, after satisfying all the bounds due to gravitino overproduction, we can have big parameter space for $T_{re}$ which is well inside {\it Planck'18} 1-$\sigma$ bound on $n_s$.
\end{abstract}
\maketitle
%%%%%%%%%%%%%%%%%%%%%
\section {Introduction}
Cosmological inflation is an era of rapid exponential expansion of the universe which is necessary to solve the initial condition problems (e.g.~horizon, flatness problem). It was quickly perceived that inflation not only solves the initial condition problems but also essential in realising the structural formation of the Universe due to the fluctuations of the inflaton field \cite{1,2}. There are numerous models of inflation proposed in literature (cf. \cite{1, 3}) since the idea was first established by Alan Guth\cite{4}. For the earlier seminal works on inflation reader is suggested to go through \cite{5,6, 7, 8, 9}.

With the recent advancement in observational cosmology, Cosmic Microwave Background(CMB) experiments such as the Wilkinson Microwave Anisotropy Probe ({\it WMAP})~\cite{10}, Planck mission~\cite{11} have constrained the inflationary observables quite stringently. In particular, the Planck 2015 inflation analysis \cite{11} has  ruled out many popular models of inflation. In 2018, the final results by the Planck mission is reported in \cite{12,13} which has constrained the inflationary models even more.
 
 As gravity plays a crucial role in cosmology, consideration of local supersymmetry i.e, Supergravity models in particle physics could be very much relevant in the context of inflation. It has been found that Starobinsky type plateau inflation potential \cite{14} could satisfy low tensor to scalar ratio($r$) and such potential could be achieved in Supergravity with appropriate choice of K\"ahler potential of no-scale form \cite{15,16} where the quadratic term in the scalar potential is suppressed. There are both $F$ term and $D$ term scalar potential in Supergravity models and either one of them could play the role of inflationary potential\cite{17,18,19,20}. However, in general, with $F$ term there is so called $\eta$ problem resulting in lack of required slow roll necessary for inflation\cite{20,21}.~Here, we shall consider the case where $D$ term mainly plays the role of inflation, while $F$ term is sub-dominant. With appropriate choice of no-scale K\"ahler potential, superpotential  and the gauge kinetic function, it is possible to obtain a Starobinsky like plateau inflation with Higgs and sneutrino scalar fields \cite{22,23}. However, although supergravity inspired power law plateau inflation potential could give small $r$, but as found in \cite{24}, the number of e-folding($N_e$) during inflation is much lesser than the required number of e-foldings to match the observations.~Very recently from $F$ term scalar potential with kinetic term for the inflation field in approximate canonical form, low tensor scalar ratio has been obtained \cite{25}. However, in this work with appropriate choice of K\"ahler potential and  superpotential with up and down type Higgs scalar fields, we have obtained scalar potential mainly dominated by $D$  term along with canonical kinetic term for the inflation field. For such potential it has been shown that low $r$ as well as justified number of e-foldings could be achievable apart from satisfying other CMB observables.
 
The rest of the paper is organised as follows. In section \ref{sec:sugra}, we will discuss the basic formalism of Supergravity and $F$ and $D$ term scalar potential. In section \ref{sec:inf}, the inflationary observables are calculated using the potential proposed in  section \ref{sec:sugra} and are compared with the recent observational bounds. Then in section \ref{sec:reh}, we have analysed the reheating era after the end of inflation and reported the bounds on the reheating temperature ($T_{re}$) as well as reheating number of e-foldings ($N_{re}$) for different equation of states during reheating ($w_{re}$). Finally in section \ref{sec:con} we have drawn the conclusion from our analysis.
%%%%%%%%%%%%%%%%%%%%%%%%%%%%%%%%%%%%%%%%%%%%%%%%%%%%%%%%%%%%%%%%%%%%%%%%%%%%%%%
\section{Inflationary Potential From Supergravity}
\label{sec:sugra}
$N=1$ supersymmetry has lots of resemblance with the Standard Model of particle Physics  as far as the matter fields corresponding to one of the supersymmetric partners are concerned. This can be the effective theory at low energy which is hierarchically much smaller than the Planck mass. In that case, low energy dynamics could be expected to be governed by $N=1$ supergravity theory. In $D=4$ and $N=1$ supergravity models the tree level scalar potential $V$ has contributions from $F$ term and $D$ term and expressed as:
\bea
V= V_F + V_D~.
\eea
$V_F$ is determined in terms of superpotential $W$ and the K\"ahler potential $K$ which are functions of chiral scalar superfields $\phi_i$ and $\phi_i^*$ and is written as: (Reduced Planck mass $M_P=1$)
\bea
V_F=e^G  \left[ \frac{\partial G}{\partial \phi_i } K_{ij*}
\frac{\partial G}{\partial \phi_j^* }-3\right]~,
\label{eqn1}
\eea
where  K\"ahler function, $G= K + \ln (W) + \ln (W^*)$, $K$ and $W$ are  K\"ahler potential and superpotential respectively.  $K_{ij*}$ is   the inverse of the K\"ahler metric
\bea
K^{ij*} = \frac{\partial^2   K}{\partial \phi_i \partial \phi_j^*}.
\label{eqn2}
\eea
 Potential $V_D$ depends on gauge symmetry and is related to gauge kinetic function. The $D$ term potential is written as: 
\bea
V_D  = \frac{1}{2} \sum [Re[f_{ab}]]^{-1}D^a D^b~,
\label{eqn3}
\eea
where $D^a=  -g^a \frac{\partial G}{\partial \phi_k} (T^a)^l_k \phi_l$, and $T^a$ is the group generator, $g^a$ is the corresponding gauge coupling  and $f_{ab}$ is the holomorphic function of superfield $\phi_i$. The kinetic energy term for the scalar fields in the Lagrangian is obtained from 
\bea
\frac{1}{\sqrt{-g}}{\cal L} _{kinetic} = K^{ij*} D_{\mu} \phi_i D_{\nu}\phi_j^* g^{\mu \nu}~,
\label{eqn4}
\eea
where $D_{\mu}$ is the gauge covariant derivative.
Action of complex scalar field minimally coupled to gravity consists of kinetic and potential parts and is written as:

\begin{equation}
S=  \int d^4 x \sqrt{-g} \left[ \frac{1}{\sqrt{-g}}  \mathcal{L}_{kinetic} - V(\phi _i, \phi^*_i) \right] \nonumber \; .
\end{equation}
Following \cite{15,26,16,27} we consider K\"ahler potential as:
 \bea
 K  = -3  \ln\left[1-  \frac{1}{3 } \left(  H_u^\dag H_u  +  
H_d^\dag  H_d  \right) \right]
 \eea
where $H_u$ and $H_d$ are up and down type Higgs scalars. Such construction corresponds to no scale supergravity \cite{26,27}, as the supersymmetry breaking scale remains undetermined at the tree level and the scale may be set by considering perturbative corrections. Construction of K\"ahler potential for more than single chiral superfields was particularly considered in \cite{16,17} and their stable de Sitter vacua were discussed in \cite{28}.  We have considered the following holomorphic term in the superpotential:

 \bea
 W =   \mu H_u. H_d   + e^{-(c + a_1 H_u^T.H_u + a_2 H_d^T.H_d)}
\eea
in which we will ignore the term $\mu H_u. H_d  $ in our subsequent discussion because $\mu$ parameter is considered to be relatively very small at high energy scale of inflation but becomes significant near supersymmetry breaking scale. There is no natural scale for the parameter $\mu$. We consider it of the order of electroweak scale. But the second exponential term involving up-type and down-type Higgs scalar field will be important in our analysis. We have considered $H_u$ and $H_d$ to be real.  The value of dimensionless parameter $a_1$ and $a_2$ are almost of equal magnitude and will be fitted from our analysis. The Yukawa interaction terms associated with masses of lepton and quarks which are also  holomorphic, have not been considered in the superpotential due to smallness of Yukawa couplings.
The up and down type Higgs scalars $H_u$ and $H_d$ are written as 
\begin{align}
H_u &= \begin{pmatrix}
\phi_u^+ \\ 
\phi_u^0
\end{pmatrix} ;  &
H_d &= \begin{pmatrix}
\phi_d^0 \\ 
\phi_d^-
\end{pmatrix}  \; .
\end{align} 
For writing the potential one may note that the $SU(2)_L$ symmetry generators are the Pauli matrices $\tau^a/2$, $U(1)_Y$ hypercharges for $H_u$ and $H_d$ are 1/2 and -1/2 respectively. At the time of inflation we consider that the neutral scalar components $\phi_u^0$ and $\phi_d^0$ will play the significant role and other components are negligible. Vaccum expectation values ($vev$) are zero for charged scalars.  We choose gauge kinetic function $f_{ab} = \delta_{ab}$ in Eq.~(\ref{eqn3}). In Eqs. (\ref{eqn1},\ref{eqn2},\ref{eqn3},\ref{eqn4}) $\phi \equiv H_{\alpha}$ where $\alpha=u,d$ and $\phi_i$ is the $i^{th}$ component of the column matrix  $H_{\alpha}$. In the expression of $D^a$ sum over $\alpha$ is implied in $H_{\alpha}$.  Ignoring the $\mu$ term in the superpotential for its smallness with respect to   the energy scale where inflation occurs, the $F$ and $D$  term for the scalar potential can be written  as:

\begin{equation}
V_{F}= -\frac{3  e^{-{2 \left(c+{a_1} {\phi_u^0}^2+{a_2} {\phi_d^0}^2\right)}} \left(-12  \left(({a_1}-1) {a_1} {\phi_u^0}^2+{a_2}^2 {\phi_d^0}^2-{a_2} {\phi_d^0}^2\right)+4 \left({a_1} {\phi_u^0}^2+{a_2} {\phi_d^0}^2\right)^2+9 \right)}{\left(-3 +{\phi_d^0}^2+{\phi_u^0}^2\right)^2}\; ;
\end{equation}
and
\begin{equation}
V_{D}= \frac{\left({g_1}^2+{g_2}^2\right) \left(3  \left((1-2 {a_1}) {\phi_u^0}^2+(2 {a_2}-1) {\phi_d^0}^2\right)-2 \left({\phi_d^0}^2+{\phi_u^0}^2\right) \left({a_2} {\phi_d^0}^2-{a_1} {\phi_u^0}^2\right)\right)^2}{\left(6 -2 \left({\phi_d^0}^2+{\phi_u^0}^2\right)\right)^2}.
\end{equation}
\noindent

For unequal $\phi_u^0 $ and  $\phi_d^0 $  if we write $\phi_u^0 = \phi_d^0/k_1 = \phi $, where $k_1$ is some constant then although the kinetic term for the field $\phi$  is not in its canonical form, but the kinetic term can be canonicalized for a different field $\varphi$ which could be related with the field $\phi $  with suitable transformation. However, the potential $V$ will not be in appropriate form for satisfying slow roll conditions as required for inflation. To get suitable inflation potential we are required to consider $k_1=1$. Then, $\phi_u^0$=$\phi_d^0$= $\phi$. In that case, the kinetic term as follows from Eq.~(\ref{eqn4}), can be written as:
 
\bea
{\cal L}_{kinetic}  = \frac{18 }{{\left(3  -2 {\phi}^2\right)}^2}{\partial^\mu \phi\;\partial_\mu \phi}~,
\label{eqn11}
\eea
which is not in its canonical form. To relate with observational data we have to consider  field  for which the kinetic term can be written in its canonical form. 
%Using Eq.~(\ref{eqn11}) and (\ref{eqn12}) one can relate $\phi$ field  to canonical
Using the following transformation:

\bea
\phi = \sqrt{\frac{3}{2}} \; \tanh \left(\frac{\varphi}{\sqrt{3}}\right)
\eea
one gets the canonical kinetic term of the field $\varphi$. The potential $V_F$ and $V_D$ can be written in terms of field $\varphi$ as:

\bea
V_{F} &= &-3  e^{-2 c-3 (a_1+a_2) \tanh^2 \left(\frac{\varphi}{\sqrt{3}}\right)}  \cosh^4 \left( \frac{\varphi}{\sqrt{3} } \right) \times  \nonumber \\  && \left( 1-2 (-a_1+a^2_1+(-1+a_2)a_2) \tanh^2 \left( \frac{\varphi}{\sqrt{3} } \right) +(a_1+a_2)^2 \tanh^4\left(\frac{\varphi}{\sqrt{3} } \right) \right)
\eea 

\begin{equation}
V_{D}= \frac{9}{4}  ({a_1}-{a_2})^2 \left({g_1}^2+{g_2}^2\right) \tanh ^4\left(\frac{\varphi }{\sqrt{3} }\right)
\label{eqn15}
\end{equation}
\noindent
 where the full potential  $V= V_{F}+V_{D}$.   For dimensionless parameters $a_1$ and $a_2$ of almost equal order of magnitude and greater than 1, $V_F$  is exponentially suppressed with respect to $V_D$  and the complete potential $V$ is dominated by $D$ term of the full potential. The specific choices of $a_{1}$, $a_{2}$ and $c$ for fitting with the observed data are mentioned later. However, one may note here, for  $a_{1}=a_{2}$,  $V_D$ vanishes and only $V_F$ remains in the potential $V$. For suitable inflationary potential one has to consider  $a_{1}$ and $a_{2}$ to be unequal and $c$ is to be chosen around $16$ or above.  The potential satisfies the slow-roll condition required for inflation for the canonical inflation field $\varphi$. It has the property of having even slower roll than other plateau type power law potential because of the presence of $\tanh$ terms which varies more slowly at higher values of the inflation field.  We have checked that the requirement of suitable inflation potential satisfying slow roll condition, indicates that the up type neutral scalar and the down type neutral scalar fields, both should be  of  equal strength at the time of inflation.
\section{Inflationary Observables}
 Considering two gauge couplings  $g_1 \sim g_2 \sim 0.65$ at high energy scale.  The slow roll parameters $\epsilon$ , $\eta$ and $\xi$ are defined as \cite{29, 30}:
\begin{equation}
    \epsilon =\frac{1}{2}  \left(\frac{{V\;'}}{V}\right)^2~, 
   ~~~~~~\eta = \frac{{V\;''}}{V}\;~~ ,~~~~~~~ \xi = \frac{V' V'''}{V^2}
   \label{eps}
\end{equation}
%%%%%%%%%%%%%%%%%%%%
Here prime denotes the derivative with respect to $\varphi$ as usual.
The amount of inflation is described in terms of number of e-folds during the inflationary epoch and is given by: 
%%%%%%%%%%%%%%%%%%%%%

\begin{equation}
N_e = \ln \left(\frac{a_{e}}{a_i}\right) ~\simeq~ \int_{t_i}^{t_{e}} H dt = \int_{\varphi_i}^{\varphi_{e}} \frac{H}{\dot{\varphi}}~ d\varphi ~\simeq ~ {\int_{\varphi _e}^{\varphi _i} \frac{V}{{V\;'}} \, d\varphi }~
\label{efold}
\end{equation}

\begin{figure}[H]
%\label{N_e_nsr}
  \centering
  \subfloat[]{\includegraphics[height= 7cm, width=8cm]{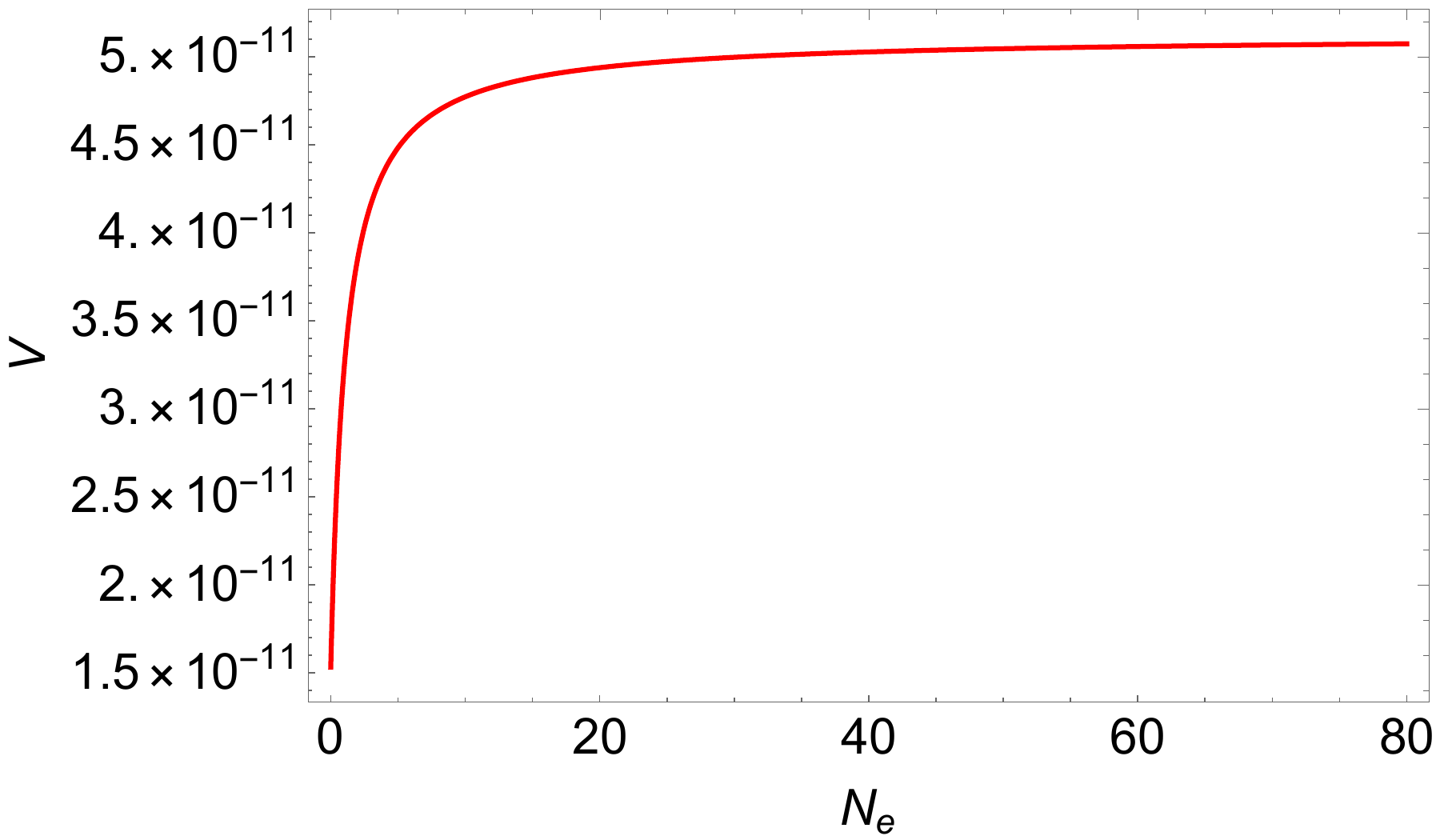}}
  \hfill
  \subfloat[]{\includegraphics[height= 7cm, width=8cm]{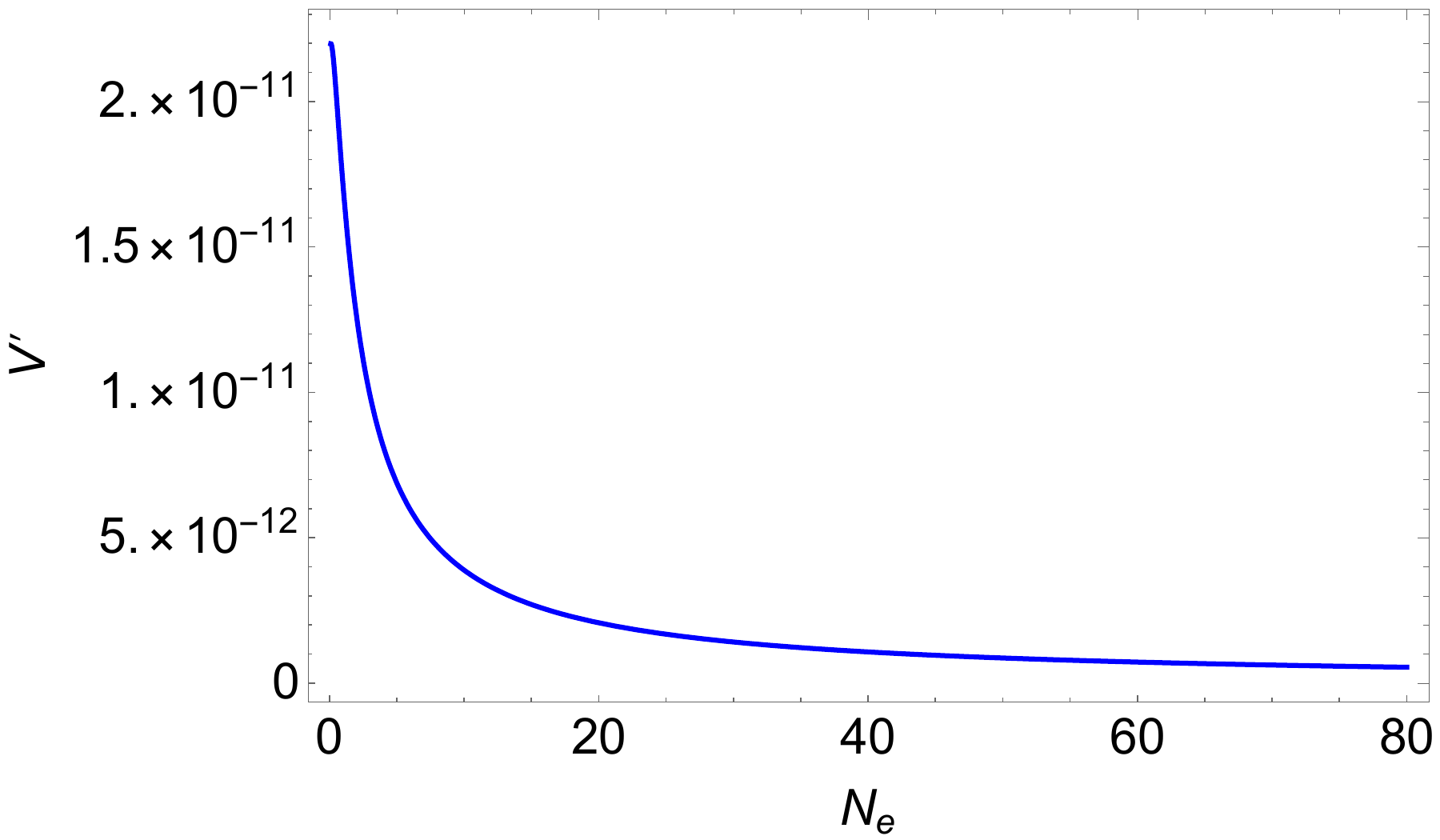}}
\caption{\small{Plots of $V$ and $V'$ as a function of $N_e$ during inflation with the suitable choices of $a_1$ and $a_2$, which are mentioned later.  }}
\label{plotvv'}
\end{figure}

\label{sec:inf}
\noindent 
where $\varphi_{e}$  denotes the end of inflation which can be calculated using the end of inflation condition ( $\epsilon = 1$) and $\varphi_{i}$ is the value of inflaton field at the time of horizon exit $\varphi_{i}= 0.8660\operatorname{arcosh}(3.41565+5.3333 N_e)$.  The potential $V$ and its derivative  $V'$, both are positive over the entire range of inflation as shown in Fig. \ref{plotvv'} and $N_e = 0$ at the end of inflation. 
 The inflationary observables- scalar spectral index $n_s$, tensor to scalar ratio ($r$) and running of the scalar spectral index ($\frac{d n_s}{d \ln k}$), are defined respectively as:
 \begin{equation}
n_s = 1 - 6 \epsilon + 2 \eta~,~~r = 16\epsilon~,~~\frac{dn_s}{d \ln k}(\equiv\alpha) = 16 \epsilon \eta - 24 \epsilon^2 - 2 \xi
\label{infobs}
\end{equation}
The amplitude of scalar perturbation is defined as:
\begin{equation}
{A_s}= \frac{1}{24 \pi^2} \left(\frac{V}{\epsilon}\right)
\label{amplitude}
\end{equation}
Using Eq. (\ref{efold}) we can write various inflationary observables -  scalar spectral index ($n_s$), tensor to scalar ratio ($r$), running of the scalar spectral index ($\frac{d n_s}{d \ln k}$) and amplitude of scalar perturbation in terms of no. of e-folds~($N_e$) as given below:
\begin{equation}
{r} = 170.667\operatorname{csch}\left(\operatorname{arcosh}(3.41565+5.3333 N_{e})\right)^2
\label{rne}
\end{equation}
\begin{equation}
{n_s} = 1-32\operatorname{csch}\left(\operatorname{arcosh}(3.41565+5.3333 N_{e})\right)^2-5.3333\operatorname{sech}\left(0.5\operatorname  {arcosh}(3.41565+5.3333 N_{e})\right)^2
\label{nsne}
\end{equation}
\begin{equation}
{A_s} = 8.1076\times 10^{-14} \operatorname{sinh}\left(0.5
\operatorname{arcosh}(3.41565+5.3333 N_{e})\right)^4 \operatorname{tanh}\left(0.5 \operatorname  {arcosh}(3.41565+5.3333 N_{e})\right)^2
\label{asne}
\end{equation}
\begin{figure}[H]
%\label{N_e_nsr}
  \centering
  \subfloat[]{\includegraphics[height= 5cm, width=5.5cm]{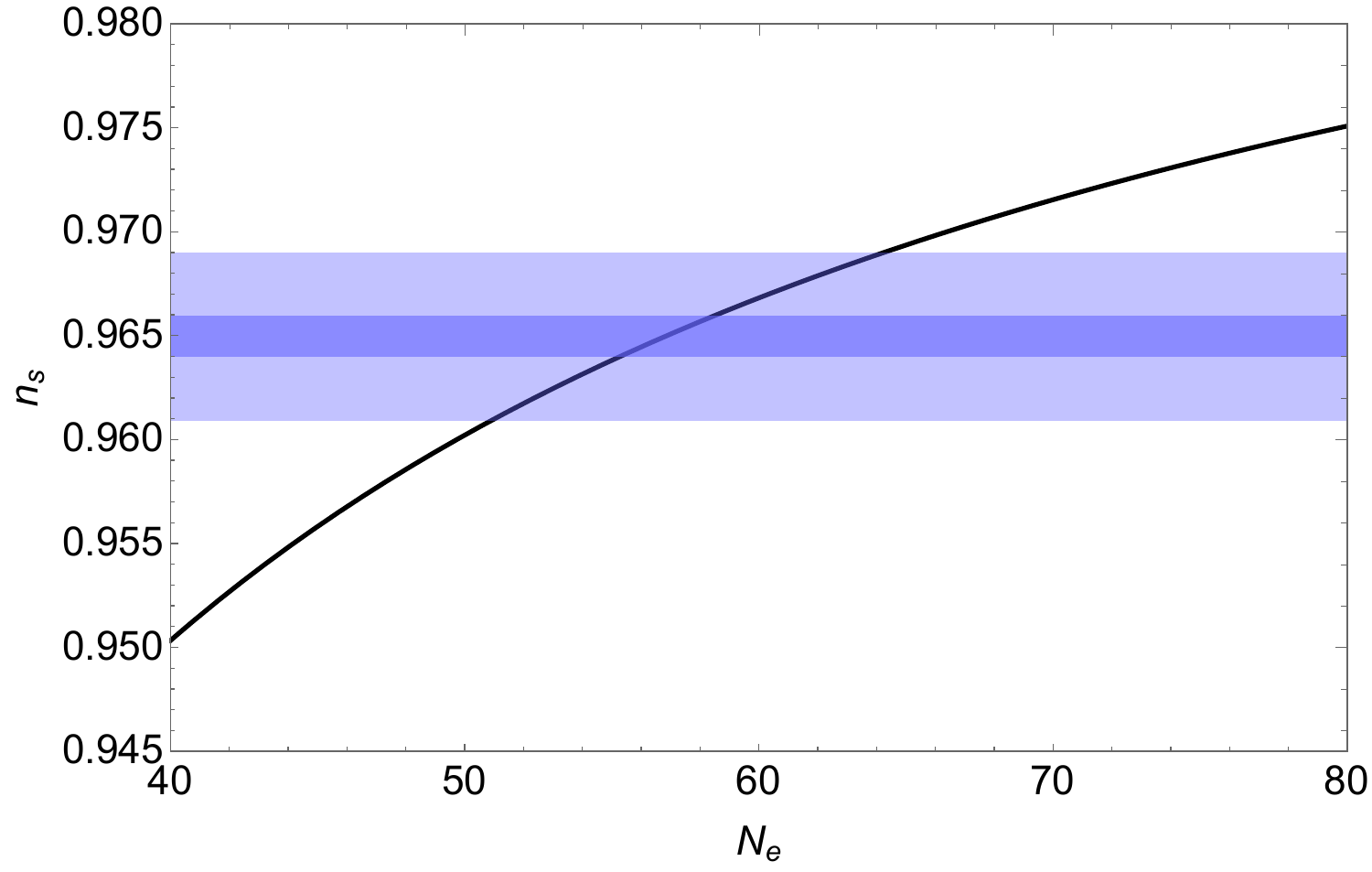}\label{f1a}}
  \hfill
  \subfloat[]{\includegraphics[height= 5cm, width=5.5cm]{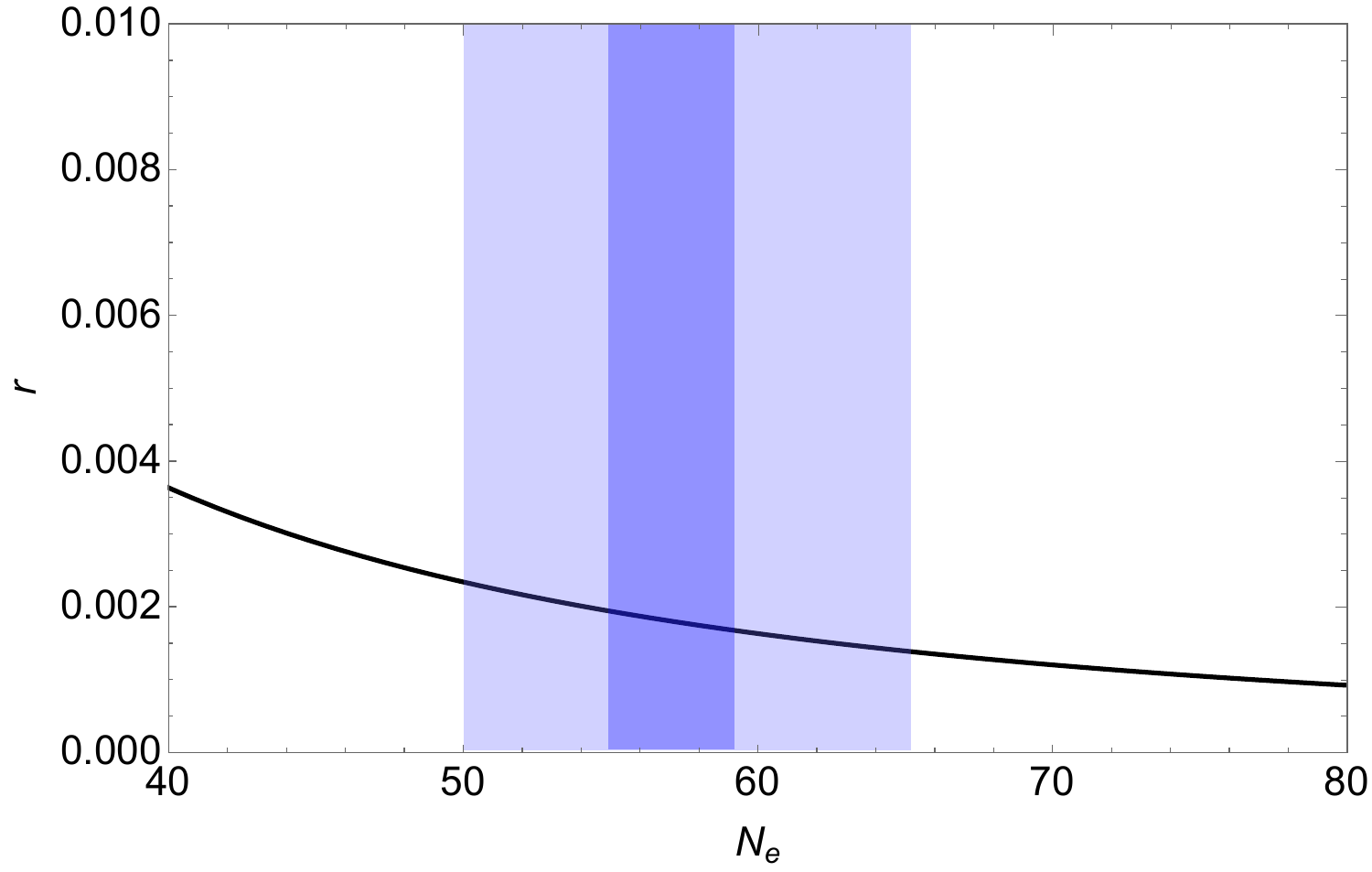}\label{f1b}}
  \hfill
  \subfloat[]{\includegraphics[height= 5cm, width=5.5cm]{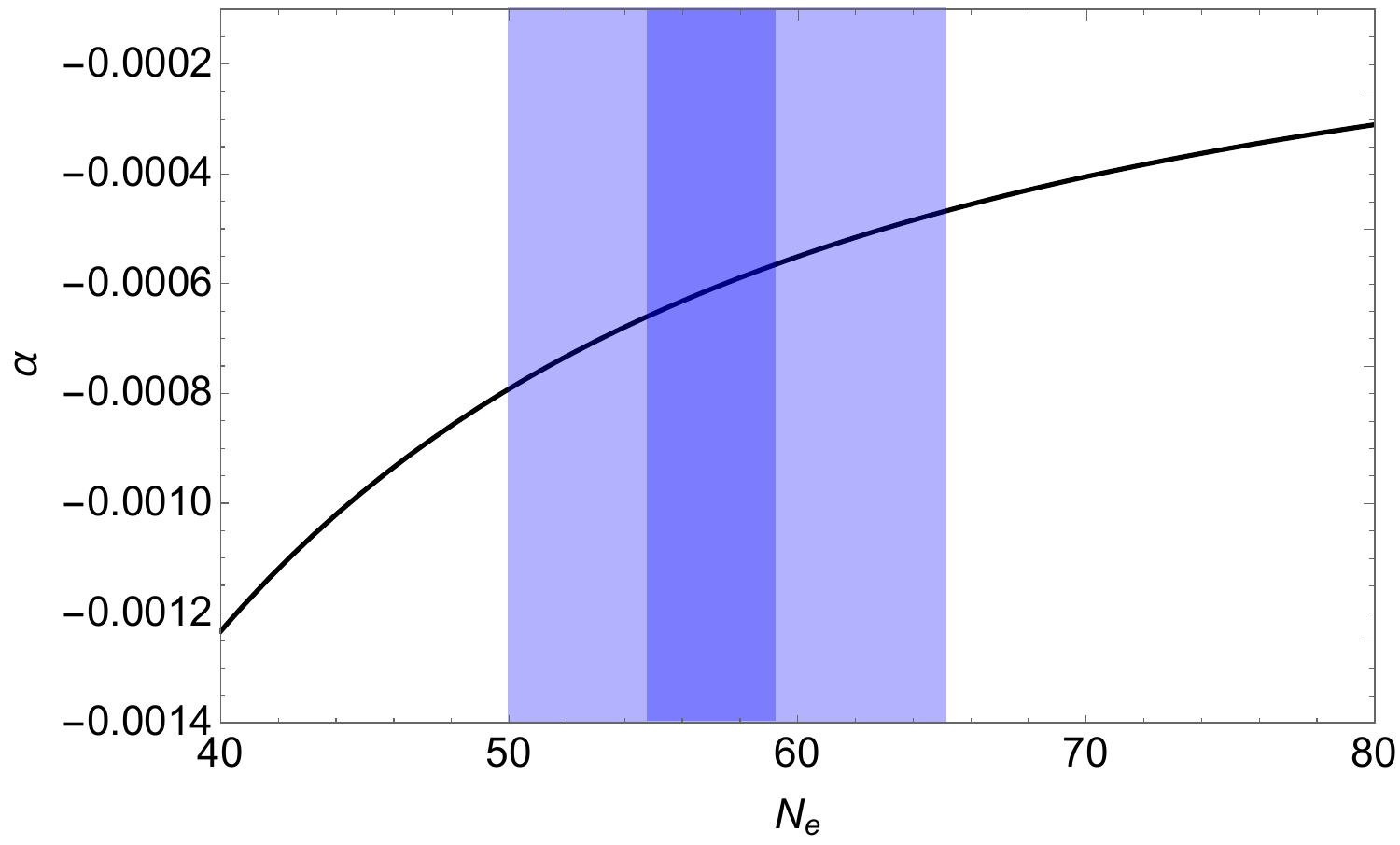}\label{f1c}}
\caption{\small{Plots of $n_s$,$r$ and $\alpha$ as a function $N_e$ respectively in \ref{f1a},\ref{f1b}, \ref{f1c}. The light blue shaded region corresponds to the 1-$\sigma$ bounds on $n_s$ from {\it Planck'18}. The deep blue shaded region corresponds to the 1-$\sigma$ bounds of future CMB observations \cite{31, 32} using the same central value for $n_s$ in \ref{f1a}. In \ref{f1b} and \ref{f1c} the bounds on $n_s$ is transferred to the bounds on $N_e$.}}
\label{plot1}
\end{figure}

%%%%%%%%%%%%%%%%%%%%%%%%%%%%%%%%
\begin{align}
\alpha = &\Big[-14.2222 - 42.6667\operatorname{csch}(0.5 \operatorname{arcosh}(3.41565+5.3333 N_e))^2  -21.3333\operatorname{csch}(0.5 \operatorname{arcosh}(3.41565+5.3333 N_e))^4\Big]\nonumber\\ 
&\times\operatorname{sech}(0.5 \operatorname{arcosh}(3.41565+5.3333 N_e))^4
\label{alphane1}
\end{align}

%%%%%%%%%%%%%%%%%%%%%%%%%%%%%%%%
The analysis is done so that  the amplitude remains consistent with the observational value measured at the pivot scale ($k$) of $0.05 \;\mbox{Mpc}^{-1}$($A_s(k_0)= 2.0989\times10^{-9}$). We have checked through our numerical analysis that one can obtain very good fit to various observed data as discussed  for the following choices of the two dimensionless parameters $a_1=7.0$ and $a_2=7.00000519$.
%For the  potential mentioned in (\ref{eqn15}) it is very difficult to solve  Eq.~(\ref{efold}) analytically. So here, we use the numerical approach, by varying the e-fold ($N_e$) over a wide range and doing the necessary cubic fitting. The choice of cubic fitting is due to the value of the norm of residuals associated with it. Cubic fitting gives significantly less value of the norm of residuals with respect to the value associated with the quadratic fitting. On the other hand, in case of quartic fitting there is no significant improvement. One can establish the relation of $n_s, r, \alpha$ in terms of $N_e$ as follows: 
 %%%%%%%%%%%%%%%%%%%%%%%%%%%%%%%%

The variations of $n_s$, $r$, $\alpha$ with $N_e$ are shown in Fig.~\ref{plot1} along with the constraints on the observables from the latest observations as mentioned in the Figure caption. The different values of $n_s$, $r$, $\alpha$  along with $A_s$ are given in Table~\ref{tab}. corresponding to different values of $N_{e}.$

\begin{center}
\begin{table*}[ht]
\begin{center}
\begin{tabular}{|l|l |r| l| l| }
\hline
\multicolumn{1}{|c|}{ {$N_e$}} & \multicolumn{1}{c|}{ $r$ } & \multicolumn{1}{c|}{ $n_s$ }& \multicolumn{1}{c|}{$A_s$}&\multicolumn{1}{c|}{$\alpha$}\\
\hline
\,\,\,\,\,\,$50$\,\,\,\,& \,\,$0.0023397$\,\, &\,\,  $0.960213$\,\,& \,\,$1.45676\times10^{-9}$&\,\,\,\,\,\,\,\,$-0.0007914$\,\,\,\,
\\
\,\,\,\,\,\,$60$\,\,\,\,& \,\,$0.0016316$\,\, &\,\, $0.966814$\,\,& \,\,$2.09402\times10^{-9}$&\,\,\,\,\,\,\,\,$-0.0005506$\,\,
\\

~~~$65$\,\,\,\,& \,\,$0.0013925$\,\, &\,\, $ 0.969357$\,\,& \,\,$2.45589\times10^{-9}$&\,\,\,\,\,\,\,\,$-0.0004694$\,\,

 \\
 \,\,\,\,\,\,$70$\,\,\,\,& \,\,$0.0012024$\,\, &\,\, $ 0.971537$\,\,& \,\,$2.84659\times10^{-9}$&\,\,\,\,\,\,\,\,$-0.0004050$\,\,
 \\
\hline
\end{tabular}
\end{center}
\caption{For $N_e= 50, \; 60, \; 65,\; 70$, the values of various inflationary parameters for the potential in (\ref{eqn15}).}
\label{tab}
\end{table*}
\end{center}
%%%%%%%%%%%%%%%%%%%%%%%%%
%%%%%%%%%%%%%%%%%%%%%%%%%%%%%%%%%%%%%%%%%%%%%%%%%%%%%%%%%%%%%%%%
\section{Reheating Parameters}
\label{sec:reh}
%%%%%%%%%%%%%%%%%%%%%%%%%%
At the end of inflation (for the cold inflationary scenario), universe ends up in a super-cooled state. Thus to enter the radiation dominated era and to start the standard Big Bang Nucleosynthesis(BBN), there is an era of reheating of the universe which is required after the end of inflation \cite{33, 34, 35, 36, 37, 38, 39}. For other realisation of inflationary dynamics e.g. Warm inflation, reader is suggested to go through Ref. \cite{40, 41, 42, 43,44} where the universe can directly enter the radiation dominated era after the end of inflation. This evolution of the universe from the supercooled state to a hot, thermal and radiation dominated state can be realised either through the perturbative reheating or the parametric resonance process better known as (p)reheating (For detailed discussion reader is suggested to follow \cite{45}). In cases of potentials like ours, the process of reheating of the universe happens mostly due to the (p)reheating process during the fast roll phase right after the end of the slow roll violation. The epoch of reheating can be parameterised by $N_{re}$ (number of e-foldings during the reheating phase), $T_{re}$ (thermalisation temperature) and the equation of states during reheating ($w_{re}$) \cite{46, 47}. Without going into the actual dynamics of governing the reheating phase one can still explore these parameters indirectly.

If one consider $w_{re}$ to be constant during the reheating era then the energy density of the universe can be related with scale factor by using $\rho \propto a^{-3(1+w)}$ as:
\begin{equation}
    \frac{\rho_{end}}{\rho_{re}} = \left(\frac{a_{end}}{a_{re}} \right)^{-3(1+w_{re})},
    \label{re1}
\end{equation}
where subscript $end$ indicates the end of inflation and $re$ indicates the end of reheating era. Replacing $\rho_{end}$ by $(3/2) V_{end}$
\begin{equation}
    N_{re} = \frac{1}{3(1+w_{re})} \ln \left(\frac{\rho_{end}}{\rho_{re}} \right)= \frac{1}{3(1+w_{re})} \ln \left(\frac{3}{2}\frac{V_{end}}{\rho_{re}} \right),
    \label{re2}
\end{equation}
The density and temperature are related as:
\begin{equation}
\rho_{re} = \frac{\pi^2}{30} g_{re} T_{re}^4.
\label{re3}
\end{equation}
Here $g_{re}$ is the number of relativistic species at the end of reheating.

Using (\ref{re2}) and (\ref{re3}) and following  \cite{48, 49, 50}, one can establish the relation between $T_{re}$ and $N_{re}$ :
\begin{equation}
N_{re} = \frac{1}{3(1+w_{re})} \ln \left(\frac{30 \cdot \frac{3}{2}  V_{end}}{\pi^2 g_{re} T_{re}^4 } \right)
\label{re4}
\end{equation}
Considering that the entropy is conserved from the reheating epoch till today, we can write  
\begin{equation}
T_{re}= T_0 \left(\frac{a_0}{a_{re}} \right) \left(\frac{43}{11 g_{re}} \right)^{\frac{1}{3}}=T_0 \left(\frac{a_0}{a_{eq}} \right) e^{N_{RD}} \left(\frac{43}{11 g_{re}} \right)^{\frac{1}{3}},
\label{re5}
\end{equation}
where $N_{RD}$ is the number of e-folds during radiation era and $e^{-N_{RD}}\equiv a_{re}/a_{eq}$. The ratio $a_{0}/a_{eq}$ can be formulated as 
\begin{equation}
\frac{a_0}{a_{eq}} = \frac{a_0 H_{k}}{k} e^{-N_{k}} e^{- N_{re}} e^{- N_{RD}}\
\label{re6}
\end{equation}

From the relation $k_{}=a_{k} H_{k}$ and using the Eq.~(\ref{re4}), (\ref{re5}) and (\ref{re6}), assuming $w_{re} \neq \frac{1}{3}$ and $g_{re} \approx 226$ (degrees of freedom in a supersymmetric scenario), we can compute the expression for $N_{re}$
\begin{equation}
N_{re}= \frac{4}{ (1-3w_{re} )}   \left[61.488  - \ln \left(\frac{ V_{end}^{\frac{1}{4}}}{ H_{k} } \right)  - N_{k}   \right]
\label{re7}
\end{equation}
Here we have used Planck's pivot  ($k$) of order $0.05 \; \mbox{Mpc}^{-1}$. In a similar way we can calculate $T_{re}$:
\begin{equation}
T_{re}= \left[ \left(\frac{43}{11 g_{re}} \right)^{\frac{1}{3}}    \frac{a_0 T_0}{k_{}} H_{k} e^{- N_{k}} \left[\frac{3^2 \cdot 5 V_{end}}{\pi^2 g_{re}} \right]^{- \frac{1}{3(1 + w_{re})}}  \right]^{\frac{3(1+ w_{re})}{3 w_{re} -1}}.
\label{re8}
\end{equation}
To evaluate $N_{re}$ and $T_{re}$ first one need to calculate the $H_{k}$, $N_{k}$ and $V_{end}$ for the given potential. Using the  definition of tensor to scalar ratio one can write
\begin{equation}
{H_k}=\sqrt{\frac{1}{2} \pi ^2 A_{s}  r}.
\label{Hk}
\end{equation}
Keeping $A_s(k_0)= 2.0989\times10^{-9}$ one can write $H_k$ in terms of $N_e$ as: 
\begin{equation}
{H_k}= 1.3295\times 10^{-3} \sqrt{\operatorname{csch(arccosh(3.41565+5.3333 N_e))^2}}
\end{equation}
From Eq.(\ref{re7}) and (\ref{re8}) we can see that both the Eqs. are the function of $H_k$ and from Eq.(\ref{Hk}) we know that $H_k$ is the function of tensor to scalar ratio. From observation one can see that there is no lower bound on the $r$, so to get the correct bound on reheating temperature  we can define $H_k$ in terms of spectral index. But from Eqs. (\ref{rne}) and (\ref{nsne}) it is not possible to write $H_k$ in terms of $n_s$. To deal with this problem, we use the numerical approach, by varying the e-fold ($N_e$) over a wide range and doing the necessary cubic fitting. The choice of cubic fitting is due to the value of the norm of residuals associated with it. Cubic fitting gives significantly less value of the norm of residuals with respect to the value associated with the quadratic fitting. On the other hand, in case of quartic fitting there is no significant improvement. One can establish the relation of $n_s, r$ in terms of $N_e$ as follows: 
\begin{equation}
    {n_{s}}=1.60372 \times{10^{-7}} N_e^3-4.08878 \times{10^{-5}} N_e^2 +0.003737 N_e+0.855481
    \label{nsn}
\end{equation}
%%%%%%%%%%%%%%%%%%%%%%%%%%%%%
\begin{equation}
 {r}=-3.46071\times{10}^{-8} {N_e^3}+8.32424\times{10}^{-6} {N_e^2}-0.0006813 N_e+0.019927
 \label{rn}
\end{equation}
%%%%%%%%%%%%%%%%%%%%%%%%%%%%%
Using Eqs.~ (\ref{Hk}), (\ref{nsn}) and (\ref{rn}) one can establish the relation between $n_{s}$ and $H_{k}$  as:  
 \begin{align}
H_{k}= \frac{\pi}{\sqrt{2}} \Big(  A_s \Big(0.106-0.107 n_s - 7.779\times10^{-42} \sqrt{1.832\times10^{80}+n_s(-3.755\times10^{80}+1.923\times10^{80} n_{s})} \nonumber\\ +\frac{1}{555648 n_s-542431+4.006\times10^{-35}\sqrt{1.832\times10^{80}+n_s(-3.755\times10^{80}+1.923\times10^{80} n_s)}} \nonumber\\ -\frac{3.597\times10^{52}}{(-3.632\times10^{86}+3.721\times10^{86} n_s +2.683\times10^{46}\sqrt{1.832\times10^{80}+n_s(-3.755\times10^{80}+1.923\times10^{80} n_s)})^{2/3}}\nonumber\\- \frac{1.081\times10^{25}}{(-3.632\times10^{86}+3.721\times10^{86} n_s +2.683\times10^{46}\sqrt{1.832\times10^{80}+n_s(-3.755\times10^{80}+1.923\times10^{80} n_s)})^{1/3}} \nonumber\\ +8.184\times10^{-32}(-3.632\times10^{86}+3.721\times10^{86}n_s+2.683\times10^{46}\sqrt{1.832\times10^{80}+n_s(-3.755\times10^{80}+1.923\times10^{80} n_s)})^{1/3} \nonumber\\-2.058\times10^{-60}(-3.632\times10^{86}+3.721\times10^{86}n_s+2.683\times10^{46}\sqrt{1.832\times10^{80}+n_s(-3.755\times10^{80}+1.923\times10^{80} n_s)}~)^{2/3}\Big)\Big)^{1/2}
\label{Hkns}
\end{align}

\begin{figure}[H]
%\label{N_e_nsr}
\centering
\includegraphics[height=6.3cm, width=0.8\textwidth]{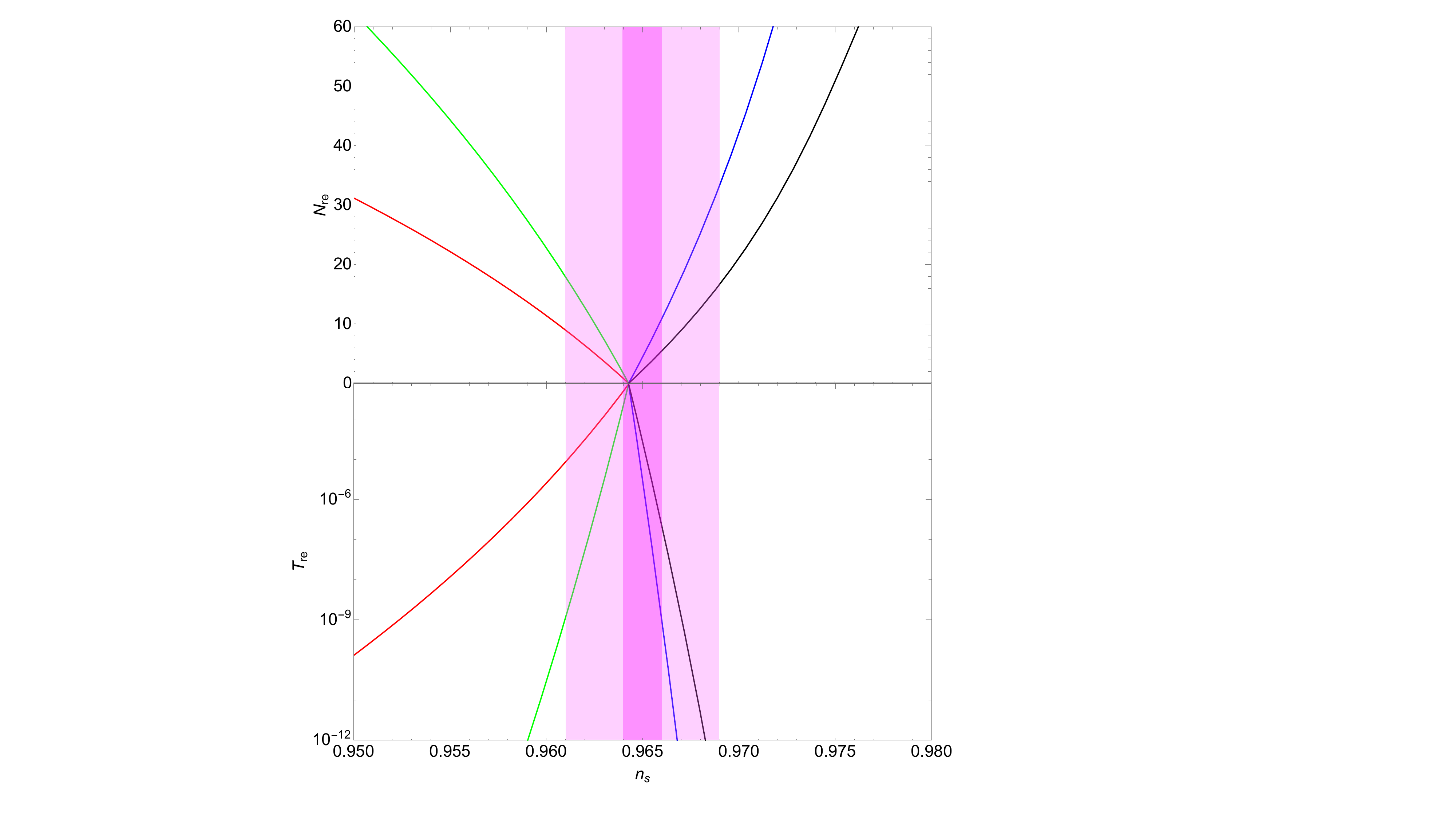}
  \hfill
\caption{\small{Plots of $N_{re}$ and $T_{re}$ as a function $n_s$ for different values of $w_{re}$ . The red line corresponds to $w_{re}= -1/3$, the green line corresponds to $w_{re}= 0$, the blue line corresponds to $w_{re}= 2/3$ and finally the black line corresponds to $w_{re}= 1$. The light pink shaded region corresponds to the 1-$\sigma$ bounds on $n_s$ from {\it Planck'18}. The dark pink shaded region corresponds to the 1-$\sigma$ bounds of future CMB observations \cite{31, 32} using same central value for $n_s$.  
%The Grey shaded region in Fig \ref{f2b} represents the electroweak phase transition. Temperature below $4.7$ MeV is ruled out by the BBN.
}}
\label{plot2}
\end{figure}
Using the inflation end condition $\epsilon = 1$ , one can calculate the $V_{end}$ and then get $T_{re}$ and $N_{re}$  by using Eq.~(\ref{re7}), (\ref{re8}) and (\ref{Hkns}) for different values of equation of state ($w_{re}$). In consistence with the definition of $N_e$ in Eq. (\ref{efold}), $N_{re}$ is negative. However, in  Fig. \ref{plot2}, $N_{re}$ which has been mentioned, is actually  $|N_{re}|$. The variation of $T_{re}$ and $N_{re}$ for different values of $w_{re}$ is shown in Fig.~\ref{plot2}. We would like to mention that the merging points for the $T_{re}$ plot and the $N_{re}$ plot correspond to the instant reheating scenario thus making $N_{re}=0$. 

%%%%%%%%%%%%%%%%%%%%%%%%%%%%%%%%%%%%%%%%%%%%%%%%%%%%%%%%%%%%%%%%%%%%%%%%%%%%%%% 
\section{Conclusion}
\label{sec:con}
In this work we have introduced a particular Supergravity model of inflation in which the D-term of the potential plays the dominant role. The requirement of slow roll condition for the inflation potential implies that  the up type neutral scalar and the down type neutral scalar are with equal field strength at the time of inflation. If this relationship holds down to the electroweak scale for the cooresponding $vev$ values of these fields, then it will indicate higher SUSY breaking scale around 100 TeV \cite{51,52}. This could have some implications at low energy phenomenology that we would like to explore in future. The potential responsible for inflation in our case, is coming for a completely canonical Lagrangian.  We have shown, for our case, all the inflationary observables well satisfies the {\it Planck'18} bounds. Also we would like to emphasize that for our inflationary potential low tensor to scalar ratio ($r$) in the order $\mathcal{O}(10^{-3})$ is achievable. 

We have also studied the reheating era and calculated the related reheating temperature and the related number of e-foldings. Obviously, when one is studying the inflation and reheating in a supersymmetric model, gravitino overproduction problem needs to be dealt with to have a successful BBN. However, the presence of the gravitino leads to serious cosmological problems depending on its mass and nature. If the gravitino is unstable and has a mass $m_{3/2}$ in the range of $\mathcal{O}(100)$ GeV to $\mathcal{O}(10)$ TeV, then it can completely destroy the notion of successful BBN. To achieve a successful phase of BBN, $T_{re}$ has to be less than $10^7-10^8$ GeV. On the other hand, if the gravitino is as light as $m_{3/2} < \mathcal{O}(10)$ GeV and it is stable (that is, the lightest supersymmetric particle (LSP)), the reheating temperature should satisfy $[T_{re}\leq \mathcal{O}(10^7) \; \mbox{GeV} \; (m_{3/2}/1 \; \mbox{GeV})]$ for $m_{3/2}\leq 100$ keV for the gravitino density not to exceed the observed dark matter density \cite{53}. On that note, we would like to comment, for our model with $w_{re}=2/3$ and $w_{re}=1$, after satisfying all the bounds due to gravitino overproduction, we can have big parameter space for $T_{re}$ which is well inside {\it Planck'18} 1-$\sigma$ bound on $n_s$.

A reconstructed study of inflationary potential \cite{54,55} in a Supergravity framework and its effects on reheating could be an interesting work that we would like to explore in the future. Also, a parameter estimation using the Monte Carlo Moarkov Chain(MCMC) approach could give us the better understanding of the model as in the case initiated for string motivated models in \cite{56, 57}, which could be used to explore the Supergravity parameter space indirectly. We hope to come back to these issues in recent future.

 %%%%%%%%%%%%%%%%%%%%%%%%%%%%%%%%%%%%%%%%%%%%%%%%%%%%%%%%%%%%%%%%%%%%%%
 \begin{acknowledgments}
 Work of MRG is supported by Department of Science and Technology, Government of India under the Grant Agreement number IF18-PH-228 (INSPIRE Faculty Award). The authors would like to thank A.~A.~Sen, B.~R.~Dinda for useful discussions. The authors would  like to thank the anonymous referee for several highly helpful comments.
 \end{acknowledgments}
 %%%%%%%%%%%%%%%%%%%%%%%%%%%%%%%%%%%%%%%%%%%%%%%%%%%%%%%%%%%%%%%%%%%%%%%%%


\begin{thebibliography}{99}
%%%%%%%%%%%%%%%%%%%
%%%%%%%%%%
\bibitem{1}
A. R. Liddle and D. H. Lyth, {\it Cosmological Inflation and Large 
Scale Structure}, (Cambridge University Press: Cambridge, UK), (1998).
%

\bibitem{2}
E. W. Kolb and M. S. Turner, 
{\it The Early Universe}, Addison-Wesley, Menlo Park, Ca., 1990

\bibitem{3} 
  J.~Martin, C.~Ringeval and V.~Vennin,Phys.\ Dark Univ.\  {\bf 5-6}, 75 (2014); arXiv:1303.3787.
  %  
\bibitem{4}
A. H. Guth, {Phys.\ Rev.\ D {\bf 23}, 347 (1981)}.
%
\bibitem{5}
  A.~D.~Linde, {Phys.\ Lett.\  {\bf 108B}  389 (1982)}.
%
  
  \bibitem{6} 
  A.~H.~Guth and S.~Y.~Pi, {Phys.\ Rev.\ Lett.\  {\bf 49}, 1110 (1982)}.
  %
\bibitem{7} 
  A.~D.~Linde, {Phys.\ Lett.\  {\bf 129B}, 177 (1983)}.
  %
\bibitem{8} 
  P.~J.~Steinhardt and M.~S.~Turner, {Phys.\ Rev.\ D {\bf 29}, 2162 (1984)}.
  
%  
\bibitem{9} 
  A.~Albrecht and P.~J.~Steinhardt, {Phys.\ Rev.\ Lett.\  {\bf 48}, 1220 (1982)}.
  
 %
\bibitem{10} 
G. Hinshaw  \textit{et al.}., Astrophys. J. Suppl. Ser., {\bf 208}, 19 (2013); arXiv:1212.5226.
%
\bibitem{11}
P.~A.~R.~Ade  \textit{et al.},\ Astron. $\&$ \ Astrophys, {\bf 594} A20 (2016); arXiv:1502.02114.
%

\bibitem{12} 
  N.~Aghanim {\it et al.}, Astron. Astrophys. \textbf{641}, A6 (2020), erratum: Astron. Astrophys. \textbf{652}, C4 (2021); arXiv:1807.06209.



\bibitem{13} 
 Y.~Akrami {\it et al.}, Astron. Astrophys. \textbf{641}, A10 (2020); arXiv:1807.06211.
%
\bibitem{14}
A. A. Starobinsky,{Phys. Lett. B  {\bf 91}, 99 (1980)}

\bibitem{15} 
  J.~Ellis, D.~V.~Nanopoulos and K.~A.~Olive,Phys.\ Rev.\ Lett.\  {\bf 111}, 111301 (2013), Erratum: Phys.\ Rev.\ Lett.\  {\bf 111}, no. 12, 129902 (2013); arXiv:1305.1247 .
  
  
 \bibitem{16} J. R. Ellis, C. Kounnas and D. V. Nanopoulos,  Nucl. Phys. B {\bf 241}, 406 (1984)
 %%
\bibitem{17}
J.~R.~Ellis, C.~Kounnas and D.~V.~Nanopoulos,Nucl. Phys. B \textbf{247}, 373-395 (1984)
 
\bibitem{18} 
  V.~Domcke and K.~Schmitz,
  %``Unified model of D-term inflation,''
  Phys.\ Rev.\ D {\bf 95}, no. 7, 075020 (2017); arXiv:1702.02173.
  
 \bibitem{19} 
  V.~Domcke and K.~Schmitz,
  %``Inflation from High-Scale Supersymmetry Breaking,''
  Phys.\ Rev.\ D {\bf 97}, no. 11, 115025 (2018); arXiv:1712.08121. 
%%%
\bibitem{20}
  S.~Zheng, Nucl.\ Phys.\ B {\bf 919},1  (2017); arXiv:1610.00406
  
\bibitem{21}
M.~Bastero-Gil and S.~F.~King, Nucl. Phys. B \textbf{549}, 391-406 (1999); arXiv:9806477.  
  
  %
  \bibitem{22} 
  G.~K.~Chakravarty, G.~Gupta, G.~Lambiase and S.~Mohanty, Phys.\ Lett.\ B {\bf 760}, 263 (2016); arXiv:1604.02556.
  
  
  \bibitem{23} 
    G.~K.~Chakravarty, U.~K.~Dey, G.~Lambiase and S.~Mohanty,  Phys.\ Lett.\ B {\bf 763}, 501 (2016); arXiv:1607.06904.
  
 
 
\bibitem{24} 
  K.~Dimopoulos and C.~Owen, Phys.\ Rev.\ D {\bf 94}, no. 6, 063518 (2016); arXiv:1607.02469.
  
 

\bibitem{25}
G.~Germ\'an, J.~C.~Hidalgo, F.~X.~Linares Cede\~no, A.~Montiel and J.~A.~V\'azquez, Phys. Rev. D \textbf{101}, no.2, 023507 (2020); arXiv:1909.02019. 
 



 \bibitem{26} 
 E. Cremmer, S. Ferrara, C. Kounnas and D. V. Nanopoulos, Phys. Lett. B {\bf 133}, 61 (1983)


\bibitem{27}
J.~Ellis, B.~Nagaraj, D.~V.~Nanopoulos and K.~A.~Olive, JHEP \textbf{11}, 110 (2018); arXiv:1809.10114.


\bibitem{28}
J.~R.~Ellis, A.~B.~Lahanas, D.~V.~Nanopoulos and K.~Tamvakis, Phys. Lett. B \textbf{134}, 429 (1984)



\bibitem{29} 
  A.~R.~Liddle, P.~Parsons and J.~D.~Barrow, Phys.\ Rev.\ D {\bf 50}, 7222 (1994); arXiv:9408015.
 
\bibitem{30} 
  D.~Baumann; {arXiv:0907.5424}.

\bibitem{31}
 Euclid Theory Working Group Collaboration, L. Amendola et al., Living Rev.Rel. {\bf 16}, 6 (2013); arXiv:1206.1225.

 \bibitem{32}
 PRISM Collaboration Collaboration, P. Andre et al.; arXiv:1306.2259.
 
  
\bibitem{33} 
  V.~F.~Mukhanov, H.~A.~Feldman and R.~H.~Brandenberger, Phys.\ Rept.\  {\bf 215}, 203 (1992).
 
 \bibitem{34}   
 Andreas Albrecht, Paul J. Steinhardt, Michael S. Turner, and Frank Wilczek, Phys. Rev. Lett. {\bf 48}  (1982)
 

 \bibitem{35} 
  L.~Kofman, A.~D.~Linde and A.~A.~Starobinsky, Phys.\ Rev.\ Lett.\  {\bf 73}, 3195 (1994); arXiv:9405187.
  
 
\bibitem{36} 
  Y.~Shtanov, J.~H.~Traschen and R.~H.~Brandenberger, Phys.\ Rev.\ D {\bf 51}, 5438 (1995); arXiv:9407247.
  
  

\bibitem{37} 
  L.~Kofman, A.~D.~Linde and A.~A.~Starobinsky, Phys.\ Rev.\ D {\bf 56}, 3258 (1997); arXiv:9704452.
  
  

\bibitem{38} 
  B.~A.~Bassett, S.~Tsujikawa and D.~Wands, Rev.\ Mod.\ Phys.\  {\bf 78}, 537 (2006); arXiv:0507632.
  
 

\bibitem{39} 
  T.~Rehagen and G.~B.~Gelmini, JCAP {\bf 1506}, no. 06, 039 (2015); arXiv:1504.03768.
  
 

\bibitem{40} 
  A.~Berera, Phys.\ Rev.\ Lett.\ {\bf 75}, 3218 (1995); arXiv:9509049.


\bibitem{41} 
  A.~Berera and L.~Z.~Fang, Phys.\ Rev.\ Lett.\  {\bf 74}, 1912 (1995); arXiv:9501024.
  
 
  

\bibitem{42} 
  M.~Bastero-Gil et. al., Phys.\ Rev.\ Lett.\  {\bf 117}, no. 15, 151301 (2016); arXiv:1604.08838.
 
 
  
 
\bibitem{43} 
  M.~Bastero-Gil et. al., JCAP {\bf 1802}, 054 (2018); arXiv:1710.10008.
  %
  
  
  

\bibitem{44}
M.~R.~Gangopadhyay, $et,al,$ Phys. Rev. D \textbf{103}, no.4, 043505 (2021); arXiv:2011.09155.


  \bibitem{45} 
  K.~D.~Lozanov;  arXiv:1907.04402.
  
  
 
\bibitem{46} 
  J.~Martin, C.~Ringeval and V.~Vennin, Phys.\ Rev.\ Lett.\  {\bf 114}, no. 8, 081303 (2015); arXiv:1410.7958.
  
  
 
\bibitem{47} 
  R.~C.~de Freitas and S.~V.~B.~Gonçalves; arXiv:1509.08500.
  
  
 
\bibitem{48} 
  J.~L.~Cook, E.~Dimastrogiovanni, D.~A.~Easson and L.~M.~Krauss, JCAP {\bf 1504}, 047 (2015); arXiv:1502.04673.
 
  
  
\bibitem{49} 
  R.~G.~Cai, Z.~K.~Guo and S.~J.~Wang, Phys.\ Rev.\ D {\bf 92}, 063506 (2015); arXiv:1501.07743.
  
  
  
  
  \bibitem{50} 
  J.~O.~Gong, S.~Pi and G.~Leung, JCAP {\bf 1505}, no. 05, 027 (2015); arXiv:1501.03604.
  %``Probing reheating with primordial spectrum,''
 
 
\bibitem{51} 
  A.~Djouadi and J.~Quevillon, JHEP {\bf 1310}, 028 (2013); arXiv:1304.1787.
 
 
 
\bibitem{52}
A.~Djouadi, L.~Maiani, A.~Polosa, J.~Quevillon and V.~Riquer, JHEP \textbf{06}, 168 (2015); arXiv:1502.05653.
  
  

  
  \bibitem{53}
  M.~Kawasaki, F.~Takahashi and and T.~T.~Yanagida, Phys.Rev. D {\bf 74} ,043519  (2006); arXiv:0605297.
  %
  \bibitem{54}
 W.~H.~Kinney,\ Phys.\ Rev.{\bf D66} (2002) 083508; arXiv:0206032.
 %
 
\bibitem{55}
S.~Bhattacharya, K.~Das and M.~R.~Gangopadhyay, Class. Quant. Grav. \textbf{37}, no.21, 215009 (2020); arXiv:1908.02542.

%
 \bibitem{56}
 S.~Bhattacharya et. al., Phys.Rev. D { \bf 97}, no.12, 123533  (2018); arXiv:1711.04807.
%


\bibitem{57}
S.~Bhattacharya, K.~Dutta, M.~R.~Gangopadhyay, A.~Maharana and K.~Singh, Phys. Rev. D \textbf{102}, 123531 (2020); arXiv:2003.05969.


\end{thebibliography}
  \end{document}